\documentclass[]{aastex701}

\begin{document}

\title{A stellar stream around the spiral galaxy Messier 61 in Rubin First Look imaging}

\author[orcid=0000-0003-2473-0369,gname=Aaron,sname=Romanowsky]{Aaron J.\ Romanowsky}
\affiliation{Department of Physics \& Astronomy, San Jos\'e State University, One Washington Square, San Jose, CA 95192, USA}
\affiliation{Department of Astronomy \& Astrophysics, University of California Santa Cruz, 1156 High Street, Santa Cruz, CA 95064, USA}
\email[show]{aaron.romanowsky@sjsu.edu}  

\author[orcid=0000-0003-3835-2231,gname=David, sname=Mart\'inez-Delgado]{David Mart\'inez-Delgado} 
\affiliation{Centro de Estudios de F\'isica del Cosmos de Arag\'on (CEFCA), Unidad Asociada al CSIC, Plaza San Juan 1, 44001 Teruel, Spain}
\affiliation{ARAID Foundation, Avda. de Ranillas, 1-D, E-50018 Zaragoza, Spain}
\email{dmartinez@cefca.es}

\author[orcid=0000-0003-2536-5092,gname=Giuseppe, sname=Donatiello]{Giuseppe Donatiello} 
\affiliation{UAI - Unione Astrofili Italiani /P.I. Sezione Nazionale di Ricerca Profondo Cielo, 72024 Oria, Italy}
\email{donatiello_giuseppe@libero.it}

\author[orcid=0000-0003-1808-1753,gname=Juan, sname=Mir\'o-Carretero]{Juan Mir\'o-Carretero} 
\affiliation{Departamento de F\'isica de la Tierra y Astrof\'isica, Universidad Complutense de Madrid, E-28040 Madrid, Spain}
\affiliation{Leiden Observatory, Leiden University, Gorlaeus Building at Einsteinweg 55, NL-2333 CC Leiden, The Netherlands}
\email{miro.juan@gmail.com}

\author[orcid=0000-0003-1250-8314,gname=Seppo, sname=Laine]{Seppo Laine} 
\affiliation{IPAC, Mail Code 314-6, Caltech, 1200 E. California Blvd., Pasadena, CA 91125, USA}
\email{seppo@ipac.caltech.edu}

\begin{abstract}

We present the first stellar stream discovered with the Vera C.\ Rubin Observatory,
around spiral galaxy M61 (NGC~4303) in Virgo First Look imaging.
The stream is narrow, radially-oriented in projection, and $\sim$~50~kpc long.
It has $g$-band surface brightness (SB) $\mu_g \sim 28$~AB~mag~arcsec$^{-2}$,
color $g-z \sim 1.0$, and stellar mass $M_\star \sim 2\times 10^8 M_\odot$.
This dwarf galaxy interaction may have provoked
the M61 starburst, and foreshadows the bounty of
accretion features expected in the ten-year Rubin Legacy Survey of Space and Time (LSST).

\end{abstract}

\section{Introduction} 

Giant spiral galaxies like the Milky Way (MW) 
constantly accrete dwarf galaxies that disrupt into stellar streams,
as hallmarks of the hierarchical universe,
useful for testing
galaxy formation and dark matter theories \citep{Nibauer25}.
MW halo streams are discovered by resolved 
stars, e.g., the dramatic 
Sgr-dwarf disruption
\citep{Majewski03}.
Beyond the Local Group, diffuse light is used,
e.g., with DECaLS
to $\mu_g \sim 28.5$~mag~arcsec$^{-2}$ 
\citep{Martinez-Delgado23}.
LSST will provide a major advance in studying
nearby galaxy halos,
given the anticipated depth ($\mu_g \sim$~30--31~mag~arcsec$^{-2}$)
and sky-area surveyed \citep{Laine18,Martin22}.

During main camera (LSSTCam; \citealt{LSSTCam,LSSTpipeline}) commissioning,
Rubin imaged $\sim 25$~deg$^2$
of Virgo in $ugri$,
with five-year-LSST depth\footnote{\url{https://rubinobservatory.org/news/rubin-first-look/cosmic-treasure-chest}}.
These First Look images were 
released in June 2025,
with well-studied 
galaxies revealing more exquisite detail than 
seen before.  
The image-processing 
preserved extended low-surface-brightness (LSB) features -- a 
non-trivial feat for mosaiced imagers.

One dramatic novelty 
is a long, narrow stellar stream 
extending Northward from the MW-like galaxy M61
(noticed by G.~Donatiello using an ED127mm f/9 refractor in 2020\footnote{\url{https://flic.kr/p/2kc7Sjr}}).
The face-on spiral disk is 
studied in PHANGS, with
$\sim 180$~km~s$^{-1}$ rotation-velocity \citep{Lang20},
$M_\star = 4\times 10^{10} M_\odot$ \citep{Lee22},
and $\sim$~10~Myr-old nuclear starburst 
\citep{Dametto19}.
We assume Virgo-cluster 16.7-Mpc distance.

\begin{figure*}[ht!]
\includegraphics[width=\textwidth]{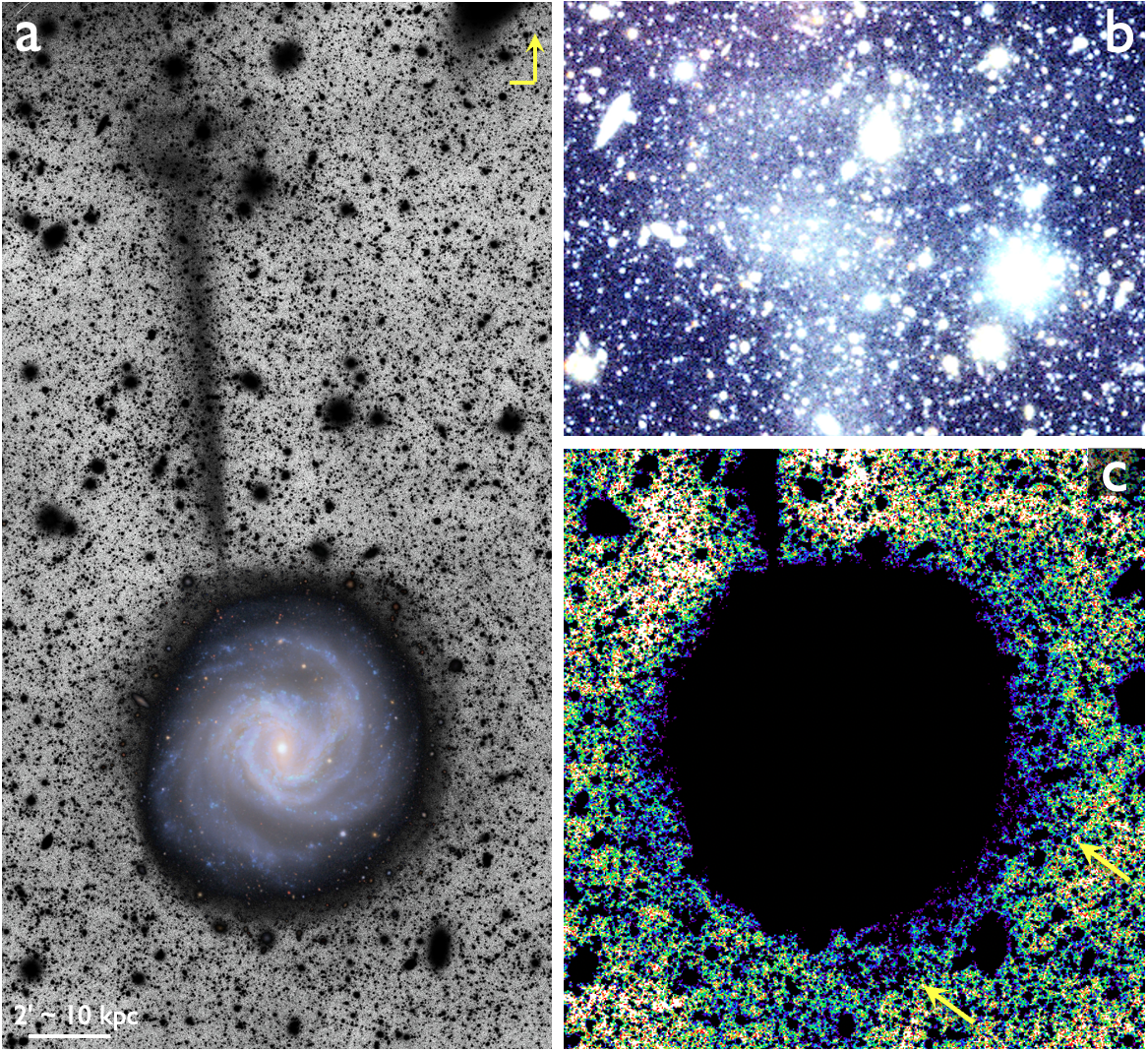}
\caption{M61 in Rubin.
(a):
50~kpc stream 
extends
Northward.
(b): Zoom-in on plume North-end,
showing complex structure.
(c): Image-stretch highlighting lower-SB features.  
Apparent plumes
$\sim$~20--30~kpc out on the disk-periphery
({\it arrows})
will require confirmation with the full dataset.
Image credit: RubinObs/NOIRLab/SLAC/NSF/DOE/AURA.
\label{fig:general}}
\end{figure*}

\section{Stream characterization} \label{sec:image}

The First Look imaging is not fully science-ready,
and we focus on 
morphology (insensitive to
photometric calibration).
For flexible visualization, we used
a CDS {\tt FITS} 
RGB-image-cube\footnote{\url{https://alasky.cds.unistra.fr/Rubin/CDS_P_Rubin_FirstLook/}}, 
from NOIRLAB
``Cosmic Treasure Chest'' {\tt TIFF} images\footnote{\url{https://noirlab.edu/public/images/noirlab2521a/}}. 

Figure~\ref{fig:general} shows the stream,
which starts at the disk-edge
($\sim$~20 kpc from center)
with $\sim$~2~kpc width, and continues very straight
for 44~kpc, 
widening to $\sim$~4~kpc, where it
terminates in a small plume
($\sim 9\times4$~kpc), with fainter extension
northward by $\sim 5$~kpc.

The stream is barely visible in DECaLS DR10 \citep{Dey19}, 
which we use for photometry
(pending the full Rubin release).
Using 
{\tt Gnuastro}\footnote{\url{https://www.gnu.org/software/gnuastro/}},
we performed secondary background-subtraction, masked contaminants, and
measured aperture-photometry
along the stream
\citep{Miro-Carretero24}.
The $g$-band SB declines from 27.2~mag~arcsec$^{-1}$ near the disk,
to 28.6~mag~arcsec$^{-2}$ toward its end, with mean of 27.9~mag~arcsec$^{-1}$.
The colors are 
$(g-r)_0 = 0.70$ and $(g-z)_0 = 1.00$
(uncertainties $\sim 0.1$~mag),
like a quenched dwarf \citep{Paudel23}.
We estimate total luminosity $L_g \sim 9 \times 10^7 L_{g,\odot}$
-- similar to Sgr
\citep{Niederste-Ostholt10}.
We use color 
to estimate stream 
$M_\star \sim 2\times10^8 M_\odot$ \citep{delosReyes25}.

The stream's orbital distance 
resembles 
the Sgr stream
\citep{Bonaca25} although the M61 stream may be narrower
\citep{Ramos22}, and at an earlier stage of disruption (with debris spanning much less than a full orbital phase).
Given an infall halo mass of $\sim 8\times10^{10} M_\odot$ 
expected from its stellar mass \citep{Wechsler18},
the stream progenitor galaxy could be responsible for the
bar formation, starburst, and active galactic nucleus in M61
\citep{Iles22},
reminiscent of the Sgr impact on the MW
\citep{Ruiz-Lara20}.
Further insights could be obtained using 
chemodynamical tracers and models of the stream \citep{Foster14}.

It is remarkable that 
the stream went long
unnoticed around a Messier galaxy. We expect a treasure trove of
substructures to be unveiled around other galaxies with future Rubin data.

\begin{acknowledgments}

We thank the Rubin team for LSB-friendly image-processing.
Supported by 
SJSU Division of Research and Innovation 
(Award 25-RSG-08-135)
and NASA under Contract No.\ 80GSFC21R0032.

\end{acknowledgments}

\facilities{Rubin:Simonyi(LSSTCam)}

\bibliography{new.ms}{}
\bibliographystyle{aasjournalv7}

\end{document}